\documentclass[aps,prl,twocolumn,floatfix]{revtex4}
\usepackage{graphicx}
\usepackage{dcolumn}
\begin{document}
\title{Failure of geometric electromagnetism in the adiabatic vector Kepler problem}
\author{J.R.~Anglin$^1$}\author{J.~Schmiedmayer$^2$}
\affiliation{$^1$Center for Ultracold Atoms, Massachusetts Institute of Technology, 
77 Massachusetts Ave., Cambridge, MA 02139\\
$^2$Institute for Experimental Physics, Heidelberg University, Heidelberg, GERMANY}

\begin{abstract}
The magnetic moment of a particle orbiting a straight current-carrying
wire may precess rapidly enough in the wire's magnetic field to justify an
adiabatic approximation, eliminating the rapid time dependence of the magnetic
moment and leaving only the particle position as a slow degree of freedom.  To
zeroth order in the adiabatic expansion, the orbits of the particle in the plane
perpendicular to the wire are Keplerian ellipses.  Higher order post-adiabatic 
corrections make the orbits precess, but recent analysis of this `vector Kepler 
problem' has shown that the effective Hamiltonian incorporating a post-adiabatic scalar 
potential (`geometric electromagnetism') fails to predict the precession correctly, 
while a heuristic alternative succeeds.  In this paper we
resolve the apparent failure of the post-adiabatic approximation, by pointing
out that the correct second-order analysis produces a third Hamiltonian, in
which geometric electromagnetism is supplemented by a tensor potential.  The heuristic 
Hamiltonian of Schmiedmayer and Scrinzi is then shown to be a canonical transformation 
of the correct adiabatic Hamiltonian, to second order.  The transformation has the important advantage of removing a $1/r^3$ singularity which is an artifact of the adiabatic approximation.
\end{abstract}

\maketitle

\section{I: Introduction}

An analogue to the Kepler problem of motion in an inverse square force can
be realized with magnetism and adiabaticity -- except that the analogue to
the planetary mass is not a scalar, but a vector component \cite{J1}.  For a particle
of mass $M$ with spin $\vec{s}$ and gyromagnetic ratio $\mu $, moving in a
magnetic field $\vec{B}(\vec{x})$, the Hamiltonian is 
\begin{equation}\label{H0}
H={\frac{|\vec{p}|^{2}}{2M}}-\mu \vec{s}\cdot \vec{B}\;.
\end{equation}
If the field $\vec{B}={\cal B}\hat{\theta}r^{-1}$ is due to a uniform
current flowing in a straight, thin wire along the $z$-axis, the $z$
component of linear momentum is conserved, and so the problem reduces to
finding the orbit of the particle in the $xy$ plane. The magnitude $s$ of 
$\vec{s}$ is fixed, and the $z$ component of total angular momentum 
$J_z=xp_{y}-yp_{x}+s_{z}$ is a constant of the motion. In realistic cases
the dimensionless ratio $\varepsilon =s/J_z$ is small.  This implies a separation of time scales, so that adiabatic methods may be applied \cite{Messiah,Berry,ShapWilc,GVP1,AhSt,L,GEM2}.

To zeroth order in $\varepsilon$ one can (as we will review below) replace $H$
with the effective Hamiltonian 
\begin{equation}
H_{AD0}={p_r^2\over2}+{p_\theta^2\over2r^2}-\frac{{\sigma}}{{r}}\;,  \label{Ham0}
\end{equation}
where we drop the $z$ direction, rescale to dimensionless units, and use
standard polar co-ordinates ${r},{\theta}$, centred on the
current-carrying wire.  In $H_{AD0}$ 
the pairs ${r},{p}_{r}$ and ${\theta},{p}_{\theta }$
are canonically conjugate, and ${\sigma}$ is simply a constant. We can
recognize $H_{AD0}$ as the Hamiltonian for Kepler's problem of
motion under an inverse square force. See Figure 1. In this vector version
of the problem, however, the analogue of the particle mass, ${\sigma}
\propto s_{\theta }$, is a vector component.  It may therefore be positive
or negative; but since the problems considered in this paper are much less
salient for unbound motion, we will assume ${\sigma}>0$.
\begin{figure}[ht]
\includegraphics[width=0.95\linewidth]{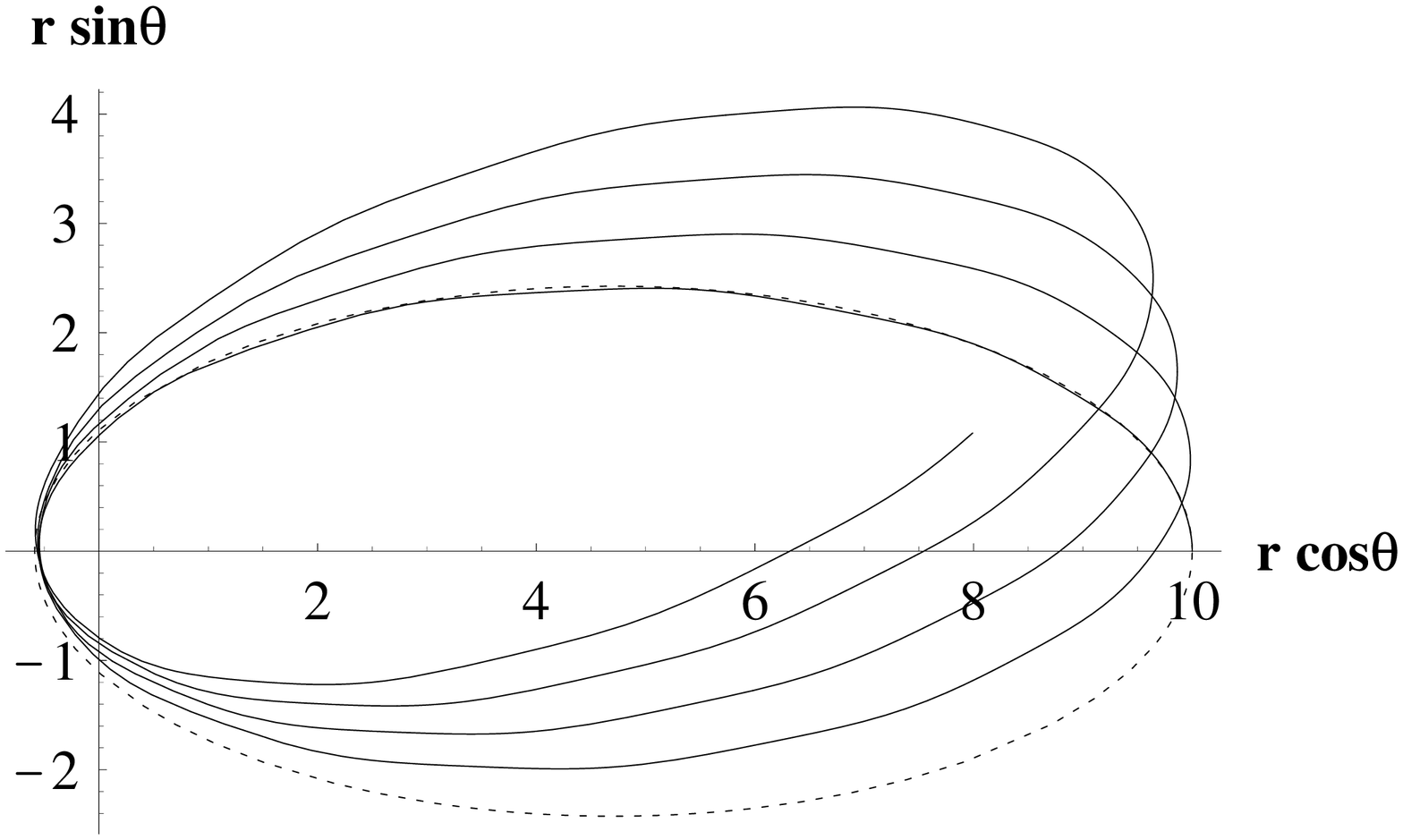}
\caption{Orbits of the magnetic particle about the current-carrying wire at the origin, 
from numerical
solution of the equations of motion (\ref{can1})-(\ref{can2})
derived from (\ref{H0}).  Initial conditions are 
$\sigma(0)=0.9$, $\alpha (0)=0$, $r(0)=10$, $p_r(0)=0$,
and $\varepsilon =s/J_z=0.2$. The
slightly polygonal appearance of the trajectory, with rounded corners
between straighter sections, is not a numerical artifact: it is the rapid
component of the particle motion. }
\end{figure}

As Figure 1 illustrates, for small $\varepsilon $ the particle orbits are
indeed well described as ellipses. Two kinds of corrections, however, are
also evident: a rapid wobble, which is straightforward to compute, and a
slow precession, which is not. Capturing effects like this slow precession
is the goal of higher order adiabatic approximations, which add to $H_{AD0}$
terms of higher order in $\varepsilon$. Classic papers on adiabatic methods 
\cite{Berry,ShapWilc,GVP1,AhSt,GEM2} prescribe for this problem the second
order effective Hamiltonian 
\begin{equation}
H_{GEM}=\frac{{p}_{r}^{2}}{2}+\frac{{p}_{\theta }^{2}+{1\over2}\varepsilon
^{2}(1-{\sigma}^{2})}{2{r}^{2}}-\frac{{\sigma}}{{r}},
\label{GEM}
\end{equation}
whose correction term can (as we will explain) be called `geometric
electromagnetism'. But see Figures 2 and 3: $H_{GEM}$ is actually worse than 
$H_{AD0}$, in that it yields precession in the wrong direction; whereas the
precession is given with impressive accuracy by a slightly different
alternative that has been identified, without derivation, by Schmiedmayer
and Scrinzi \cite{JA}: 
\begin{equation}\label{HSS}
H_{SS}=\frac{{p}_{r}^{2}}{2}+\frac{{p}_{\theta }^{2}+{1\over2}\varepsilon
^{2}(1-3{\sigma}^{2})}{2{r}^{2}}-\frac{{\sigma}}{{r}}.
\label{SS}
\end{equation}
So what has gone wrong with adiabatic theory?  And what is $H_{SS}$?
\begin{figure}[ht]
\includegraphics[width=0.95\linewidth]{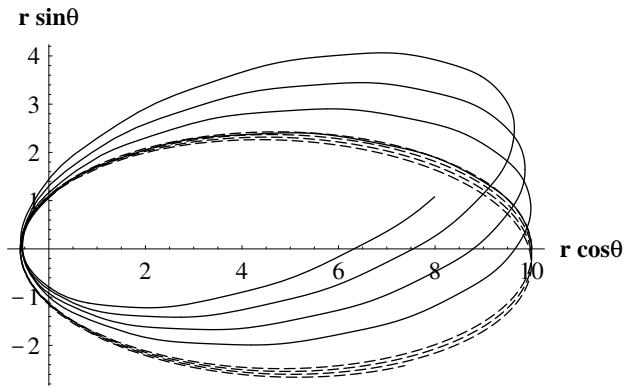}
\caption{The same as Fig.~1 (solid curve), with the geometric electromagnetism
approximation superimposed for comparison (dashed curve). The dashed curve
is the numerical solution of the canonical equations of motion
derived from the Hamiltonian $H_{GEM}$ of (\ref{GEM}), with the same initial conditions as in Fig.~1.}
\end{figure}
\begin{figure}[ht]
\includegraphics[width=0.95\linewidth]{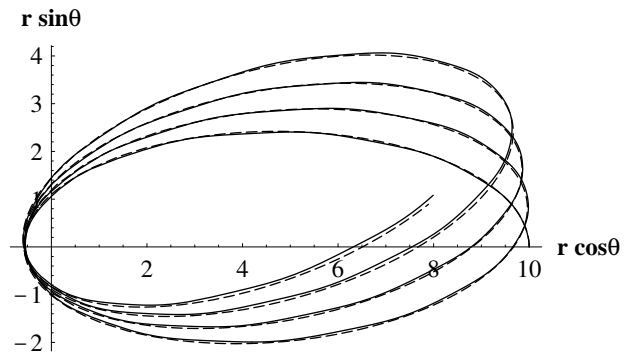}
\caption{Again the same as Fig.~1 (solid curve), with the heuristic
approximation of Schmiedmayer and Scrinzi superimposed for comparison
(dashed curve). The dashed curve is the numerical solution of
the canonical equations of motion derived from the Hamiltonian $H_{SS}$ of (\ref{HSS}),
with the same $\varepsilon$ and initial conditions as in Figs.~1 and 2.}
\end{figure}

In this paper we explain that nothing has gone wrong with adiabatic theory.
The problem is simply that a tensor potential, formally identified by
Littlejohn and Weigert \cite{L}, and proportional in our case to $r^{-3}$, 
must be included in addition to the more widely
known vector and scalar potentials of geometric electromagnetism. In the
next four Sections (II-V) we build up to a simple derivation of this term, first reviewing the zeroth order approximation, and then geometric electromagnetism.  In Section IV we discuss the subtlety of relating the initial conditions for the exact and approximate evolutions.  Then in Section V we present the tensor potential for the
classical version of our vector Kepler problem, and also provide a simple
physical explanation of the phenomenon (which is much more general).  In Section VI we
show that $H_{SS}$ is the result of a canonical transformation of the correct
second order adiabatic Hamiltonian; and we show that this transformation is particularly convenient because it removes the $r^{-3}$ singularity.   After treating the quantum
version of the problem in Section VII, we discuss our results and conclude.

\section{II: The vector Kepler problem via adiabatic approximation}

Dropping the trivial $p_z^2$ term, the Hamiltonian (\ref{H0}) may be written
\begin{equation}\label{H1}
H={p_r^2\over2M}+{L_z^2\over2Mr^2}-{{\cal B}\mu\over r}\sqrt{s^2-s_z^2}\sin(\phi-\theta)
\end{equation}
where $r e^{i\theta}=x+iy$ defines the usual polar co-ordinates for the particle's position, $p_r = p_x\cos\theta+p_y\sin\theta$, $L_z=xp_y-yp_x$.  We introduce the angle $\phi$ which is canonically conjugate to $s_z$, such that $s_x+is_y=\sqrt{s^2-s_z^2}e^{i\phi}$.  $L_z$ and $\theta$, and $p_r$ and $r$, are of course the other canonically conjugate pairs.

It is convenient to use polar spin axes, so that we have $s_\theta, s_r$ instead of $s_x, s_y$, and the canonical pair $s_\theta,\alpha$ instead of $s_z,\phi$.  Defining the new canonical momentum
\begin{equation}
s_\theta = s_y\cos\theta -s_x\sin\theta = \sqrt{s^2-s_z^2}\sin(\phi-\theta)
\end{equation}
mixes the angular co-ordinate $\theta$ with the spin sector, though.  Checking the Poisson brackets, we find that to keep the transformation canonical, we must also change the momentum conjugate to $\theta$ from $L_z$ to $J_z$.  So we must write
\begin{eqnarray}\label{CT}
s_z &=& \sqrt{s^2-s_\theta^2}\cos{\alpha}\nonumber\\
L_z &=& J_z - s_z = J_z - \sqrt{s^2-s_\theta^2}\cos\alpha\;.
\end{eqnarray}

It is also convenient to make all our variables dimensionless, by rescaling them in terms of Hamiltonian co-efficients and a reference angular momentum $J$:
\begin{eqnarray}  \label{change}
t \rightarrow {\frac{J^{3}}{M(s{\cal B}\mu )^{2}}}t  \quad && \quad
H \rightarrow {\frac{M(s{\cal B}\mu )^{2}}{J^{2}}}H  \nonumber \\
r \rightarrow {\frac{J^{2}}{sM{\cal B}\mu }} r  \quad && \quad 
p_r \rightarrow {\frac{sM{\cal B}\mu }{J}} p_r\;.\nonumber \\
J_z \rightarrow J p_\theta \quad && \quad s_\theta \rightarrow s\sigma\;.
\end{eqnarray}
These are the variables with which we will proceed.  Of course $p_\theta$ is a constant of the motion, and so we can (and do) always set the numerical value of $p_\theta\to1$ by choosing $J=J_z$; but we retain it as a canonical variable, which is needed in the equation of motion for $\theta$.  In terms of these variables the exact Hamiltonian (\ref{H1}) can be finally re-written
\begin{equation}\label{Hex}
H={p_r^2\over2}+{(p_\theta-\varepsilon\sqrt{1-\sigma^2}\cos\alpha)^2\over2r^2}-{\sigma\over r}\;,
\end{equation}
where $\varepsilon\equiv s/J$.  

As we have defined it, $\sigma$ is of order $\varepsilon^0$ in the cases we consider, which is convenient for keeping track of powers of $\varepsilon$; but the dimensionless momentum canonically conjugate to $\alpha$ would be $s_\theta/J=\varepsilon\sigma$.  This gives the equations of motion for $\alpha$ and $\sigma$ an extra factor of $\varepsilon^{-1}$: 
\begin{eqnarray}\label{can1}
\dot{\sigma} &=& -{1\over\varepsilon}{\partial H\over\partial\alpha} 
	=-(p_\theta-\varepsilon\sqrt{1-\sigma^2}\cos\alpha){\sqrt{1-\sigma^2}\over r^2}\sin\alpha
	\nonumber\\
\dot{\alpha}&=&\frac{1}{\varepsilon}\frac{\partial H}{\partial \sigma}
	=-{1\over\varepsilon r}
	+{p_\theta-\varepsilon\sqrt{1-\sigma^2}\cos\alpha\over r^2}
		{\sigma\cos\alpha\over\sqrt{1-\sigma^2}} \;.
\end{eqnarray}
Our other equations of motion are the usual canonical ones:
\begin{eqnarray}\label{can2}
\dot{r}&=&{\partial H\over\partial p_r}=p_r\nonumber\\
\dot{p}_r &=& -{\partial H\over\partial r} = 
	{(p_\theta-\varepsilon\sqrt{1-\sigma^2}\cos\alpha)^2\over r^3}-{\sigma\over r^2}\nonumber\\
\dot{\theta} &=& {\partial H\over\partial p_\theta} = 
	{p_\theta-\varepsilon\sqrt{1-\sigma^2}\cos\alpha\over r^2}\nonumber\\
\dot{p_\theta} &=& -{\partial H\over \partial\theta}=0\;.
\end{eqnarray}

We can now obtain the vector Kepler Hamiltonian (\ref{Ham0}) from (\ref{Hex}) simply by setting $\varepsilon\to0$.  In this limit, $\alpha$ evolves infinitely fast according to (\ref{can1}); but $\alpha$ plays no role in the Kepler analogy, and can simply be ignored.  On the other hand, because $\alpha$ evolves so rapidly, from the equation for $\sigma$ in (\ref{can2}) we can easily see that 
\begin{eqnarray}\label{sig1}
\sigma (t)=\bar{\sigma} - \varepsilon p_\theta{\sqrt{1-\bar{\sigma}^2}\over r}\cos\alpha + {\cal O}(\varepsilon^2)\;.
\end{eqnarray}
where $\bar{\sigma}$ is a constant.  So to zeroth order in 
$\varepsilon$, $H$ is indeed equivalent to $H_{AD0}$, as claimed.  And as Fig.~1 confirms,
the evolution of $(r,\theta)$ under $H_{AD0}$ does provide a close approximation to that under $H$, over intermediate time scales.
This is not bad; but the errors in using $H_{AD0}$ do become large for $t$ of order 
$\varepsilon^{-2}$.  We can do better than this.

\section{III: Post-adiabatic corrections: geometric electromagnetism}

To improve on our simple adiabatic approximation we must become systematic
about precisely how we are expanding in powers of $\varepsilon$, because
there is more than simple perturbation theory going on.  
The very rapid evolution of $\alpha$ is a warning of this. In fact the approximate
constancy (`adiabatic invariance') of $\sigma$, on which the Kepler analogy is based, is directly connected to the
rapidity of $\alpha$.  The fact that $\alpha $ is fast means that the
Hamiltonian varies steeply as a function of its conjugate momentum $\sigma $,
and this in turn means that, since $H$ is identically conserved under time evolution,
$\sigma$ must vary only slowly.  So the rapid variation of 
$\alpha$ is no accident. And the essential feature of our adiabatic
approximation is that as well as dropping terms whose magnitude is higher order
in $\varepsilon $, we are also dropping terms, like $\alpha$, whose
variation is on the rapid timescale $\varepsilon ^{-1}t$. 

An elegant and general treatment of post-adiabatic corrections is available in Ref. \cite{L}.  
The result that is reached is physically intuitive: to eliminate rapid evolution while preserving slow evolution, one must remove rapid components from the Hamiltonian, by averaging over the rapid angular variable which is canonically conjugate to the adiabatic invariant.  If we identify $\sigma$ as the adiabatic invariant, then $\alpha$ is the rapid variable, and so we should integrate $H$ over $\alpha$.  In many problems of this sort, it is common that rapid components of the Hamiltonian appear, as they do in (\ref{Hex}), as shifts in the positional momentum.  Thus to first order in the adiabatic small parameter (in our case, $\varepsilon$), after integrating out the rapid co-ordinate one obtains an effective Hamiltonian term linear in spatial momentum, $\vec{A}\cdot\vec{p}$ for some function $\vec{A}$.  In our case, this first order $\vec{A}$ vanishes, because the integral over $2\pi$ of $\cos\alpha$ is zero.  But in general such terms do not vanish.  They imply that a neutral particle with a magnetic moment, moving in a spatially varying magnetic field, experiences a force as though it were a charged particle moving in an effective magnetic field, whose vector potential is $\vec{A}$.  This phenomenon has therefore been named `geometric magnetism'.

At second post-adiabatic order, one typically encounters a scalar potential; in our case, integrating (\ref{Hex}) over $\alpha$ produces the scalar term
\begin{equation}
{1\over2\pi}\oint\!d\alpha\,{\varepsilon^2(1-\sigma^2)\cos^2\alpha\over2r^2}
	= \varepsilon^2{1-\sigma^2\over4r^2}
\end{equation}
which produces $H_{GEM}$ as given by (\ref{GEM}).  Continuing the analogy begun at first order, this post-adiabatic scalar potential may be compared to an effective electric potential.  The effective Hamiltonian produced by averaging $H$ over $\alpha$ can therefore be said to include geometric electromagnetism.

So why does it not work very well?  Since integrating out $\alpha$ has been quite straightforward, we might wonder whether the observed inaccuracy of $H_{GEM}$ is due simply to not going on to third or higher order in $\varepsilon$.  After all, $\varepsilon=0.2$ as in Figs.~1-3 is not really such a small parameter.  But $H$ contains no terms beyond order $\varepsilon^2$, so if the rule really is to average over $\alpha$, then it is not clear where higher order corrections could come from.  And in a case where $\varepsilon$ is truly small, so that different orders in $\varepsilon$ are quantitatively obvious, we can check (see Figs.~4 and 5) that the discrepancy between $H$ and $H_{GEM}$ is indeed a problem of lower order in $\varepsilon$.  In fact there are two independent problems, which we will discuss in turn, starting with the larger but more trivial one.  
\begin{figure}[ht]
\includegraphics[width=\linewidth]{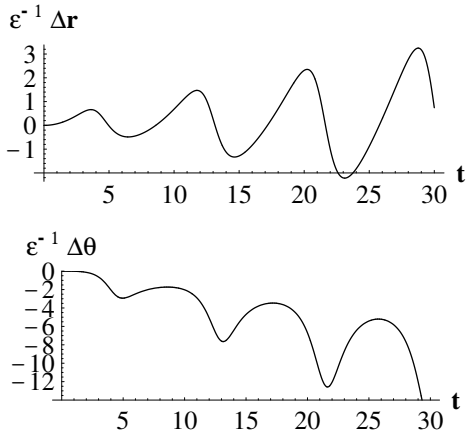}
\caption{The error $\Delta r(t) = \bar{r}-r$ and $\Delta\theta(t)= \bar{\theta}-\theta$, where $\bar{r},\bar{\theta}$ are $r,\theta$ as given by the geometric magnetism approximation. The quantities plotted are obtained by
numerically solving the same two sets of equations as in Fig.~2, but with $\varepsilon=0.01$ and $\bar{r}(0) = r(0)=1.5$
(and all other parameters the same as in the previous Figures).  The errors in the
pure geometric electromagnetism approximation are of order $\varepsilon$, rather than $\varepsilon^3$ as desired. For these parameters, however, the heuristic approximation of Schmiedmayer and Scrinzi does not fare any better: its errors are not distinguishable to the eye from those shown here.}
\end{figure}

\section{IV: Initial conditions}

It is surprising that Fig.~4 shows errors of order $\varepsilon$ in evolution under both $H_{GEM}$ and $H_{SS}$, since they agree with $H_{AD2}$ to that order; we should expect them to fail at $\mathcal{O}(\varepsilon^2)$.  In fact this expectation is correct, and the reason for the even poorer performance shown in Fig.~4 really has nothing to do with the form of the Hamiltonian.  The problem is with using the same initial conditions for the adiabatically approximate evolution as for the exact evolution.

As we can see in Eqn.~(\ref{sig1}), the system's several variables generally evolve as a superposition of slow and fast components (where the fast components may also be slowly modulated).  Adiabatic approximations seek to reproduce only the slow components.  This means that initial conditions need some consideration: the exact initial values $p_r(0)$, $\sigma(0)$, etc., are the instantaneous sums of slow and fast components of motion; but the adiabatic
approximation requires the initial values of the slow components alone.  We must therefore identify the fast components in our variables, and subtract their instantaneous values at $t=0$ from the exact initial conditions, to obtain the initial conditions that should be used with adiabatically approximate evolution.    

Fortunately it is not at all difficult to identify the fast components.  If in (\ref{can2}) we examine the equation of motion for $\theta$, for instance, we see
\begin{equation} 
\dot{\theta} = {p_\theta\over r^2} - \varepsilon\cos\alpha{\sqrt{1-\sigma^2}\over r^2}\;.
\end{equation}
Comparing this with the very rapid evolution of $\alpha$ according to (\ref{can1}), it is obvious that we must have
\begin{equation}
\theta(t) = \bar{\theta}(t) +\varepsilon^2\sin\alpha{\sqrt{1-\sigma^2}\over r}+{\cal O}(\varepsilon^3)\;,
\end{equation}
where $\bar{\theta}$ is the slow component, which does not involve $\alpha$.  Similarly we find, up to corrections of order $\varepsilon^3$,
\begin{eqnarray}\label{sols}
r(t) &=&\bar{r}(t) \nonumber \\
p_r(t) &=&\bar{p}_r(t)+\varepsilon^2 p_\theta {\sqrt{1-\sigma^2}\over r^2} \sin\alpha\nonumber\\
\sigma (t) &=&\bar{\sigma}-\varepsilon p_\theta {\frac{\sqrt{1-\sigma^{2}}\cos\alpha}{r}}+{\varepsilon^2\sigma p_\theta^2\over2r^2}\\
&& +\varepsilon^2\left( {p_\theta p_r\sqrt{1-\sigma ^2}\sin\alpha\over r}
	+{(1-\sigma ^2)\cos 2\alpha\over4r}\right)\nonumber
\end{eqnarray}
where $\bar{\sigma}$ is still exactly constant, and not just slow.  Actually, the $\varepsilon^2\sigma p_\theta^2/r^2$ term in $\sigma(t)$ is not trivial to determine; but it can be obtained fairly easily using some of the insights of the next Section, and we include it here for completeness.

So when we evolve the slow components under $H_{AD2}$, therefore, we should use $r(0)\to\bar{r}(0)$, $p_r(0)\to\bar{p}_r(0)$, etc., and $\sigma\to\bar{\sigma}$, where the barred quantities are given in terms of the exact initial conditions by
\begin{eqnarray}\label{initrans}
\bar{r}(0) &=& r(0)\nonumber\\
\bar{p}_r(0) &=& p_r(0)-\varepsilon^2p_\theta{\sqrt{1-\sigma^2(0)}\over r^2(0)}\sin\alpha(0)  \nonumber \\
\bar{\theta}(0) &=& \theta(0)- \varepsilon^2\sin\alpha(0){\sqrt{1-\sigma^2(0)}\over r(0)}\nonumber\\
\bar{\sigma} &=& \sigma(0) +\varepsilon{\sqrt{1-\sigma^{2}(0)}\cos\alpha(0)\over r(0)}
	-\varepsilon^2{\sigma(0)p_\theta^2(0)\over 2r^2(0)}\nonumber\\
&&	-\varepsilon^2{p_\theta p_r(0)\sqrt{1-\sigma ^2(0)}\sin\alpha(0)\over r(0)}\nonumber\\
&&	-\varepsilon^2{[1-\sigma^2(0)]\cos 2\alpha(0)\over4r(0)}\ .
\end{eqnarray}

We might be puzzled at this point by the fact that in replacing $\sigma(0)$ with $\bar{\sigma}$ we are actually throwing away some components that are non-constant but slow, rather than fast; this does not really seem to fit the logic behind the procedure.  Since in the adiabatic approximation $\sigma$ has to be a constant, not just a slow variable, this is clearly what we have to do.  But this point is indeed a problem: it is the problem discussed in the next Section.

Even if this subtlety did not bother us, when the adjustments (\ref{initrans}) in initial conditions and $\sigma$ are made in evolution under $H_{GEM}$ and $H_{SS}$, the first order errors of Fig.~4 are reduced, but only to second order; see Figs.~5 and 6.  Although $H_{SS}$ has no real credentials at this point, $H_{GEM}$ is supposed to be accurate at $\mathcal{O}(\varepsilon^2)$.  So we must explain the second reason for its failure.

\begin{figure}[ht]
\includegraphics[width=0.95\linewidth]{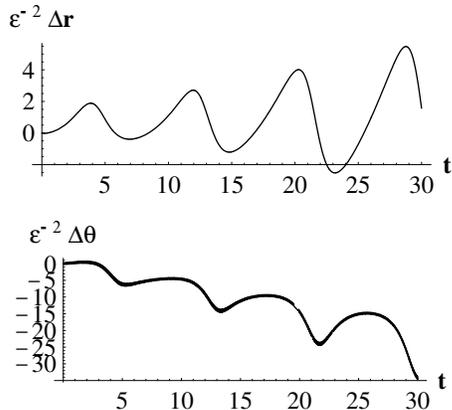}
\caption{The errors in the geometric electromagnetism approximation, with the
corrected initial conditions (\ref{initrans}).
Parameters are otherwise the same as in Fig.~4; but note the change in vertical scales.  The thickness of the $\Delta\theta$ curve reflects the rapid oscillations in the exact motion, which our horizontal scale is too coarse to resolve visibly.}
\end{figure}
\begin{figure}[ht]
\includegraphics[width=0.95\linewidth]{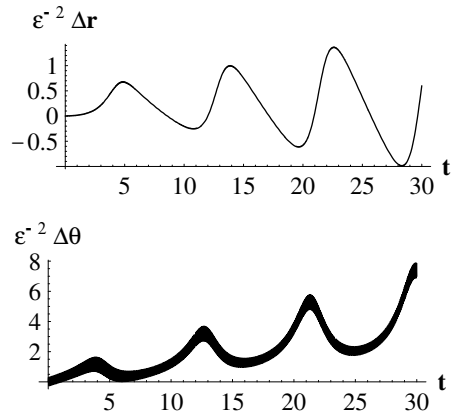}
\caption{The errors in the heuristic approximation of Schmiedmayer and Scrinzi, with the corrected initial conditions (\ref{initrans}).  Parameters are otherwise the same as in Figs.~4 and 5.  Note the vertical scale.  The thickness of the $\Delta\theta$ curve reflects the rapid oscillations in the exact motion, which our horizontal scale is too coarse to resolve visibly.}
\end{figure}

\section{V: `Geometric gravity'}

The second order problem is more profound than the first order problem of initial condtions: it is that $H_{GEM}$ is incomplete.  The scalar potential of GEM is indeed
a second order correction; but as shown by Littlejohn
and Weigert, at second order there also appears a {\it geometric tensor
potential}: the effective Hamiltonian must acquire a term of the form 
$\sum_{jk}p_{j}p_{k}g^{jk}(\vec{x})$.  In analogy with `geometric magnetism', we can think of the metric of general relativity, and call this second order
effect `geometric gravity'.  Converted to our original dimensionful units, the geometric
gravity term derived in \cite{L} is 
\begin{eqnarray}
H_{GGR} &=&-{\frac{J^{2}}{M(s\mathcal{B}\mu )^{2}}}{\frac{s\sigma }{2M^{2}\mu
B(r)}}\left| ({\vec{p}}\cdot \vec{\nabla}){\frac{\vec{B}(r)}{B(r)}}\right|
^{2}  \nonumber \\
&=&-{\frac{J^{2}s^{2}r\sigma }{2(Ms\mathcal{B}\mu )^{3}}}\left| {\frac{J_z-s_{z}}{r}}
{\frac{1}{r}}{\frac{\partial \ }{\partial \theta }}(-\sin \theta ,\cos \theta)\right| ^{2}  \nonumber \\
&\rightarrow &-{\frac{\varepsilon^2}{2}}{\frac{\sigma p_\theta^2}{r^3}}\;,  \label{GG}
\end{eqnarray}
when in this last step we return to our dimensionless variables, and drop
terms of higher order in $\varepsilon$.  This yields the
full second order adiabatic effective Hamiltonian 
\begin{eqnarray}
H_{AD2} &=&H_{GEM}+H_{GGR}  \nonumber \\
&=&{p_r^2\over2}+\frac{p_\theta^2+{\frac{1}{2}}\varepsilon ^{2}
	(1-\sigma^{2})}{2r^2}-\frac{\sigma}{r}
	-{\varepsilon^2\over2}{\sigma p_\theta^2\over r^3}\;,  \label{HAD2}
\end{eqnarray}
distinct from both $H_{SS}$ and $H_{GEM}$.

(We can already notice a point in favour $H_{AD2}$: it is numerically equal to the exact $H$ up to second order in $\varepsilon$, since with the adjusted initial conditions of (\ref{initrans}) we find $H_{AD2}(0)=H(0)+\mathcal{O}(\varepsilon^3)$, and both Hamiltonians are conserved under the evolutions they each generate.)

Where does `geometric gravity' come from?  We can see this by re-examining the equations
of motion for $\sigma$ and $\alpha$ in (\ref{can1}).  We can recall $\sigma(t)$ as given in (\ref{sols}), and then by inspection we can also find that
\begin{equation}\label{inspect}
\alpha(t) =-{\tau\over\varepsilon}
	-{\varepsilon\bar{\sigma} p_\theta\over r}{\sin{\tau\over\varepsilon}\over
		\sqrt{1-\bar{\sigma}^2}}+{\cal O}(\varepsilon^2)
\end{equation}
where $\tau(t)\equiv\int\!dt\,r^{-1}$.   These results then imply that
\begin{equation}
s_{z}\equiv s\sqrt{1-\sigma^2}\cos \alpha = s\sqrt{1-\bar{\sigma}^2}\cos{\tau\over\varepsilon} +{\varepsilon}{\bar{\sigma} p_\theta\over r}+{\cal O}(\varepsilon ^{2})\;.
\end{equation} 
So in fact the first order
correction to the purely fast zeroth order term in $\sigma _{z}$ contains
the slow term ${\varepsilon}\bar{\sigma}p_\theta /r$. Thus at first order in $\varepsilon$, 
it is no longer precisely the spin component in the $\vec{B}$ direction
which is the adiabatic invariant. The axis about which the spin precesses
rapidly tilts slightly in the $z$-direction, so that the spin component
which varies slowly is not exactly $s_{\theta}$, but rather
\begin{equation}
s_{\theta'}\equiv
{s_{\theta }+\varepsilon {p_\theta\over r} s_{z}\over\sqrt{1+\varepsilon^2{p_\theta^2\over r^2}}}\;.
\end{equation}
See Fig.~7.

\begin{figure}[ht]
\includegraphics[width=0.95\linewidth]{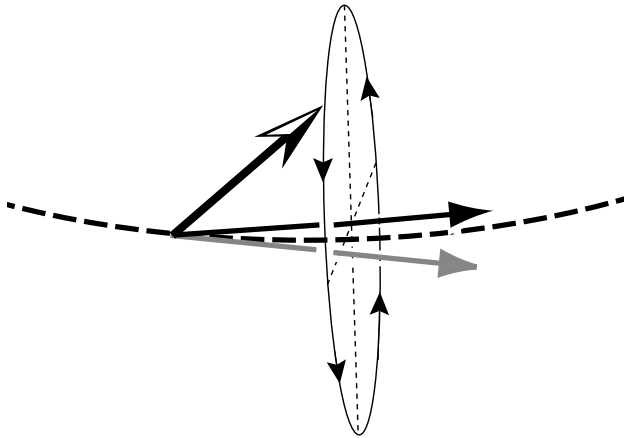}
\caption{The magnetic particle's spin (two-toned arrow) precesses rapidly around an axis (dark arrow) tilted slightly $z$-ward from the $\theta$-direction (grey arrow), which is tangential to the magnetic field line (dashed arc).  A post-adiabatic approximation that includes only geometric electromagnetism treats the spin component along the grey arrow as an adiabatic invariant.   Incorporating `geometric gravity' as well means using a more accurate adiabatic invariant, the spin component along the dark arrow.}
\end{figure}

This means that the adiabatic approximation is {\it not} exactly to average over $\alpha$,
but to average over the variable conjugate to $s_{\theta'}$.  If we now re-write the Hamiltonian $H$ in terms of $\sigma'=s_{\theta'}/s$ and its conjugate angle $\alpha'$, we find 
\begin{eqnarray}\label{Halp}
H &=& {p_r^2\over2}+{p_\theta^2+\varepsilon^2(1-\sigma^2)\cos^2\alpha\over 2r^2}
		-{\sigma'\over r}\sqrt{1+{\varepsilon^2 p_\theta^2\over r^2}}\nonumber\\
&=&	{p_r^2\over2}+{p_\theta^2+\varepsilon^2 (1-\sigma'^2)\cos^2\alpha'\over 2r^2}
	\nonumber\\
&&\qquad	-{\sigma'\over r}\left(1+{\varepsilon^2 p_\theta^2\over2r^2}\right) 
		+{\cal O}(\varepsilon^3)\;.
\end{eqnarray}
Averaging (\ref{Halp}) over $\alpha'$ now produces $H_{AD2}$, with $\sigma\to\sigma'$.
While $\sigma\to\sigma'$ is really a trivial change, since in the effective theory this quantity is simply a constant, this does resolve the concern we raised in the previous Section, immediately after Eqn.~(\ref{initrans}).  The constant component $\bar{\sigma}$ of $\sigma(t)$ is the entire slow component of $\sigma'$ (up to corrections of third order or higher); in replacing $\sigma'(0)\to\bar{\sigma}$, we are indeed discarding only fast components.

(An extra complication in this derivation of $H_{AD2}$ is that the change of variables from $\alpha,\sigma\to\alpha',\sigma'$, without also changing $p_r$ and $p_\theta$, is not canonical.  The nonvanishing Poisson brackets $[\sigma',p_r]$ and $[\sigma',\theta]$ are purely fast quantities, however, and this means that we can ignore this complication.  If we also adjusted the momenta to keep our variables exactly canonical, extra terms not shown in (\ref{Halp}) would appear when $H$ was re-written; but these extra terms would then vanish in the averaging over $\alpha'$.) 

\begin{figure}[ht]
\includegraphics[width=0.95\linewidth]{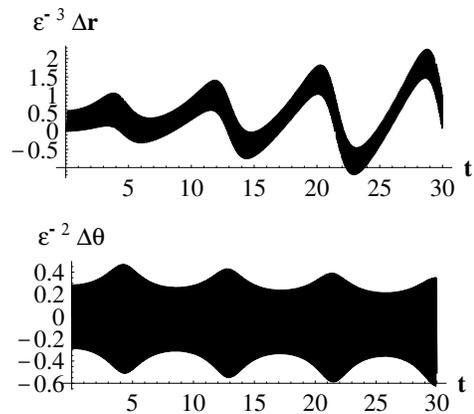}
\caption{The errors in the correct second order adiabatic approximation, evolution under $H_{AD2}$.
Parameters are again $\varepsilon=0.01$ and the initial conditions on
the exact motion are the same as in Figs.~4-6, while the initial conditions for
the $H_{AD2}$ approximation are given in terms of the exact initial conditions by
Eqns.~(\ref{initrans}). Note the change in vertical scale in comparison
with Figs.~6 and 7.  On this vertical scale the rapid component of the exact motion is quite large, but the unchanged horizontal scale is still too coarse to resolve it, producing the thick black bands.}
\end{figure}

As illustrated in Fig.~8, when $\varepsilon $ is small, $H_{AD2}$ 
with (\ref{initrans}) is accurate for the slow motion even at ${\cal O}(\varepsilon
^{2})$.  The power of the adiabatic approximation is truly
remarkable: it ignores the small, rapid `wobble' in the exact evolution, and
it tracks the slow motion very accurately over many orbits. The contrast
with Fig.~4 shows primarily the effects of correcting the initial conditions according to (\ref{initrans}), which involves a first-order change in the adiabatic invariant $\sigma$ from the instantaneous exact $\sigma(0)$.  Comparing Figs.~5 and 6, however, shows that even with the improved initial conditions, the simpler approximations $H_{GEM}$ and $H_{SS}$
are not as accurate as $H_{AD2}$ in the illustrated case.  
We now understand the failings of $H_{GEM}$; and it was not clear in the first place why one should expect $H_{SS}$ to be a good effective Hamiltonian.  But in the case shown in Figs. 1-3, $H_{SS}$ was really excellent.  Why?

\section{VI: Preferred canonical co-ordinates}

The uncanny effectiveness of $H_{SS}$, even though it does not seem to
be the correct adiabatic effective Hamiltonian $H_{AD2}$, is very
simply explained: $H_{SS}$ actually {\it is} the correct adiabatic
effective Hamiltonian, in a different set of variables. If we perform the
canonical change of variables $r\to r -{\frac{1}{2}}\varepsilon ^{2}\sigma$ in 
$H_{AD2}$, then to second order in $\varepsilon $ we obtain 
$H_{SS}$. 
Hence the heuristic evolution of Schmiedmayer and Scrinzi would disagree with the
correct adiabatic result only at third or higher order in $\varepsilon$, if in
comparing the two evolutions one also applied the small constant shift in $r$,
and used the more accurate initial conditions (\ref{initrans}).  See Fig.~9.
\begin{figure}[ht]
\includegraphics[width=0.95\linewidth]{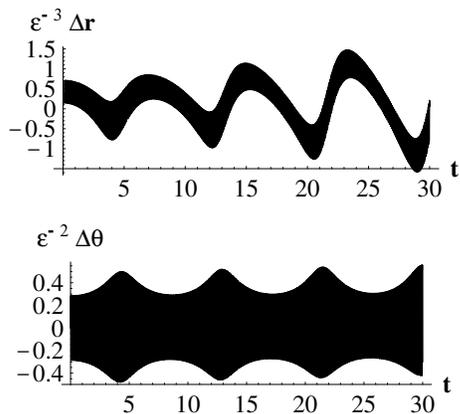}
\caption{The errors in the heuristic approximation of Schmiedmayer and Scrinzi, with the corrected initial conditions (\ref{initrans}), and the co-ordinate shift described in the text.  Evolution under $H_{SS}$ proceeds from the shifted initial condition $\bar{r}(0)=1.5+{1\over2}\varepsilon^2\bar{\sigma}$, and then $\bar{r}(t)-r(t)-{1\over2}\varepsilon^2\bar{\sigma}$ is plotted as $\Delta r$.  Note that the vertical scales are the same as in Fig.~6.}
\end{figure}

Using the trivial initial conditions instead of (\ref{initrans}) amounted to an error in $\sigma$ of $\varepsilon \sqrt{1-\sigma^2(0)}/r(0)$, which was of order 1\% in the case of Figs.~1-3, and not clearly discernible in the large-scale plot shown in our first three Figures.  In the smaller $r(0)$ case of Fig.~4, close examination revealed the error.   Taking the correct initial conditions, but failing to include the co-ordinate shift in $r$, led to the $\mathcal{O}(\varepsilon^2)$ error seen in Fig.~7.  Yet this simple co-ordinate shift is in a sense the main reason why $H_{SS}$ does so much better than $H_{GEM}$ in the strongly precessing case of Figs.~1-3: since the canonical transformation relating $H_{SS}$ and $H_{AD2}$ is a constant shift in radius, even the most naive calculation with $H_{SS}$ gives the orbital precession with full second-order post-adiabatic accuracy.

Just when everything at last seems clear, though, there is one final twist in the classical part of our story: see Fig.~10.  In some respects the canonical co-ordinates yielding $H_{SS}$
as the adiabatic effective Hamiltonian are significantly better than those yielding 
$H_{AD2}$.  For one thing, $H_{SS}$ itself is more easily solved
than $H_{AD2}$, because it is effectively just the Kepler Hamiltonian with a renormalized $p_\theta$.

More importantly, though, $H_{AD2}$ contains a term proportional to $\varepsilon^{2}r^{-3}$, while $H_{SS}$ has no terms more singular than the $r^{-2}$
centrifugal barrier. As long as $r$ remains of order $\varepsilon ^{0}$, 
the more singular term causes no problems; but for $r<\sqrt{\varepsilon }$,
the formally $\varepsilon ^{2}$ correction in $H_{AD2}$ is only smaller than the zeroth
order $\sigma /r$ term by ${\cal O}(\varepsilon )$, and the adiabatic hierarchy
will begin to break down. Of course, since the exact Hamiltonian has nothing worse than $r^{-2}$, one of the things this breakdown implies is that the $r^{-3}$ term is not really accurate when $r$ is this small.  So while the $r^{-3}$ term is quite accurate at $r\geq{\cal O}(\varepsilon^0)$, its singularity at small $r$ is spurious -- an artifact of an adiabatic expansion applied beyond its regime of validity.

In contrast, no hierarchy breakdown problems occur with 
$H_{SS}$ until $r<\varepsilon $; and even then there is no severe disaster, because the
zeroth order centrifugal barrier still dominates.  So the variables
of $H_{SS}$ are better than those of $H_{AD2}$ in that they greatly shrink
the phase space region where
the adiabatic approach fails, and also reduce the severity of the failure. 
And this advantage can translate into
significantly more accurate results, as shown in Fig.~10.  The highly eccentric orbit in this case, and the relatively large value of $\varepsilon$, mean that as the particle approaches the wire $H_{AD2}$ suffers significantly in comparison with $H_{SS}$.
\begin{figure}[ht]
\includegraphics[width=0.95\linewidth]{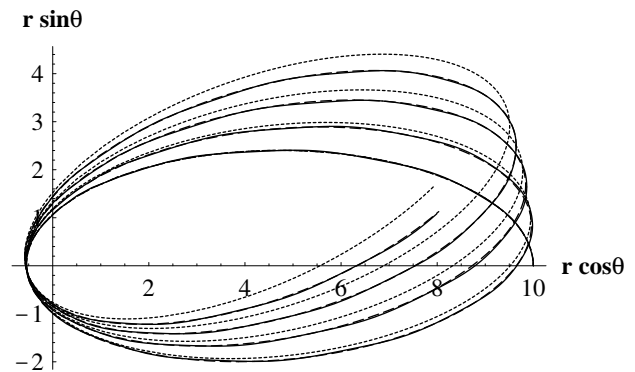}
\caption{The exact trajectory as in Fig.~1 (solid curve), with the adiabatic
trajectories as given using $H_{SS}$ (long dashes), and using 
$H_{AD2}$ (short dashes). In both approximations the
initial conditions are given by (\ref{initrans}). For $H_{SS}$ evolution, 
$r(0) = 10+\varepsilon^2\bar{\sigma}/2$, and then 
$r(t)-\varepsilon^2\bar{\sigma}/2$ is plotted as
the radius. This shift, and the improved initial conditions, make this corrected implementation of the Schmiedmayer-Scrinzi approximation even better than the naive
implementation shown in Fig.~3: it is barely distinguishable from the exact curve. Although the two dashed trajectories should formally differ only by
order $\varepsilon^3$, the breakdown of the adiabatic hierarchy as the
particle approaches the wire has a much more severe effect in the $H_{AD2}$ co-ordinates than in the Schmiedmayer-Scrinzi co-ordinates.}
\end{figure}

This raises an interesting question in adiabatic theory. When the separation
of time scales does not apply over the entire phase space, it may be that
the nonadiabatic region (in which there is no time scale separation) can be
patched into the adiabatic region (in which there is good separation of time scales) more smoothly
with some canonical coordinates than with others.  Is there a systematic way
of identifying the better co-ordinates, in general? In this paper, we can do
no more than raise this question as a subject for future investigation.

In classical mechanics this issue is not a problem if one simply avoids the
nonadiabatic region, such as by considering orbits that maintain $r>>\varepsilon$. In quantum mechanics, though, the problem is in principle
much more severe. The energy spectrum of $H_{SS}$ is well behaved, and
can even be obtained exactly (though $H_{SS}$ itself is of course only
a second order approximation); but the spectrum of $H_{AD2}$ with
positive $\sigma$ is continuously unbounded from below. One can of course
recognize the spuriousness of these negative energy states, and try to
discard them in various ways. (Semiclassical calculations, for instance,
will essentially reduce the advantages of $H_{SS}$ over $H_{AD2}$
to only those that we have noted classically.) If numerical methods are
required, however, as they may well be for more complicated problems than
the one we have been considering, a spurious singularity introduced by
adiabatic elimination might be quite a serious obstacle.

\section{VII: The quantum vector Kepler problem}

That such problems can indeed arise quantum mechanically is confirmed by
quantizing our vector Kepler problem. Keeping track of what our rescalings
have done to the commutation relations, we find the time-independent
Schr\"{o}dinger equation corresponding to (\ref{Hex}):
\begin{equation}
E\Psi ={\frac{1}{2}}\left( -{\frac{\hbar ^{2}}{J^{2}}}\Bigl[{\frac{\partial
^{2}\ }{\partial r^{2}}}+{\frac{1}{r}}{\frac{\partial \ }{\partial r}}
\Bigr]+{\frac{(1-\varepsilon \sigma _{z})^{2}}{r^{2}}}\right) \Psi -{\frac{
\sigma _{x}}{r}}\Psi \;,  \label{SE}
\end{equation}
where $\Psi =\Psi (r)$ is spinor-valued, $\sigma _{x,z}$ are the spin-$s$
generalizations of the Pauli spin matrices divided by the total spin $s$,
and $E$ is the dimensionless energy eigenvalue.  To obtain the equation in this form we have already used a $\theta$-dependent spin basis such that $s_x$ is in the local $\theta$-direction, and set $p_\theta\to 1$. 

Remembering what we learned classically about the tilting of the adiabatic invariant, we
can make a shortcut to the correct approximation at second order in $\varepsilon$
by introducing 
\begin{equation}
\psi \equiv e^{-i\varepsilon {\frac{s\sigma _{y}}{r}}}\Psi \;.  \label{psi}
\end{equation}
Since $\hbar /J$ is certainly of order $\varepsilon$, the terms that arise from the commutator between this $r$-dependent rotation and the kinetic term are of ${\cal O}(\varepsilon^3)$; but we should still keep the term linear in $\partial_r$, since $(\hbar/J)\partial_r\sim \hat{p}_r $ may be of order $\varepsilon^0$.  So up to second order, the result is
\begin{eqnarray}\label{SEBO}
E\psi &=&{\frac{1}{2}}\left( -{\frac{\hbar ^{2}}{J^{2}}}\Bigl[{\frac{\partial^{2}\ }{\partial r^{2}}}
	+{\frac{1}{r}}{\frac{\partial \ }{\partial r}}\Bigr]
	+{\frac{1+\varepsilon ^{2}\sigma _{z}^{2}}{r^{2}}}\right) \psi 
\nonumber \\
&&\qquad  -i\varepsilon{\hbar^2\over J^2}{\sigma_y\over r^2}{\partial\ \over\partial r}\psi
	-{\sigma_x\over r}\left(1+{\varepsilon^2\over2r^2}\right) \psi \;.  
\end{eqnarray}

The quantum analogue of the
classical adiabatic approximation for problems like this one 
is the Born-Oppenheimer approximation.  
In (\ref{SEBO}) we now implement the Born-Oppenheimer approximation by considering an
eigenstate of $\sigma_{x}$, with eigenvalue $\sigma$. We take the
expectation value of the Hamiltonian in this state, and solve the scalar
Schr\"{o}dinger equation 
\begin{eqnarray}
E\Phi &=&{\frac{1}{2}}\left( -{\frac{\hbar ^{2}}{J^{2}}}\Bigl[{\frac{
\partial ^{2}\ }{\partial r^{2}}}+{\frac{1}{r}}{\frac{\partial \ }{\partial r
}}\Bigr]+{\frac{1+{\frac{1}{2}}\varepsilon ^{2}({\frac{s+\hbar }{s}}-\sigma
^{2})}{r^{2}}}\right) \Phi  \nonumber \\
&&\qquad \qquad \qquad -{\frac{\sigma }{r}}\left( 1+{\frac{\varepsilon ^{2}}
{2r^{2}}}\right) \Phi \;,  \label{SSE}
\end{eqnarray}
which is clearly the quantized version of $H_{AD2}$.

If we at last define $\sqrt{r}\Phi (r)=\sqrt{r}f(r -\varepsilon
^{2}\sigma /2)$ to effect the change of variables from $H_{AD2}$ to 
$H_{SS}$ without problems from the $r^{-1}\partial _{r}$ term, we
obtain the quantized version of $H_{SS}$: 
\begin{eqnarray}
Ef &=&-{\frac{\hbar ^{2}}{2J^2}}\Bigl[{\frac{\partial ^{2}f}{\partial r
^{2}}}+{\frac{1}{r}}{\frac{\partial f}{\partial r}}\Bigr]+{\frac{1+{\frac{1}{2}}
[\varepsilon(\varepsilon+{\hbar\over J} )-3\varepsilon^{2}\sigma ^{2}]}{2r^{2}}}f  \nonumber\\
&&\qquad \qquad \qquad -{\frac{\sigma }{r}}f.  \label{SSSE}
\end{eqnarray}
In principle then we must impose boundary conditions at $r =\varepsilon
^{2}\sigma /2$ instead of at $r =0$, but since at this small $r $ we
will have $f\sim r^{J/\hbar }$, we may safely ignore this point. The
eigenvalue equation for $f$ may then be solved exactly\cite{CT}, giving us
an adiabatic approximation to order $\varepsilon ^{2}$ for the bound state
energy eigenvalues of the exact problem: 
\begin{equation}
E_{n}=-{\frac{\sigma ^{2}J^{2}}{2\hbar ^{2}}}{\frac{1}{(n-{\frac{1}{2}}+\tilde{m}_{J})^{2}}}\;,  \label{spectrum}
\end{equation}
where $n\in \lbrack 1,2,3,...]$, and the `adiabatically corrected' angular
momentum quantum number is 
\begin{equation}
\tilde{m}_{J}\equiv {\frac{1}{\hbar }}\sqrt{J^{2}+[s(s+\hbar )-3s^{2}\sigma
^{2}]/2}\;.  \label{tildeJ}
\end{equation}
Note that (\ref{spectrum}) is also obtained, to order $\varepsilon ^{2}$, by
WKB quantization of (\ref{HAD2}).

For the special case $s=\hbar /2$, (\ref{spectrum}) is not only accurate to 
${\cal O}(\varepsilon^2)$, but is actually exact \cite{s12}. This was precisely what
motivated Schmiedmayer and Scrinzi to consider their effective Hamiltonian
for general $s$. For other values of $s$ and a range of values of $J$,
numerical results in Ref.~\cite{JA} show that (\ref{spectrum}) is actually
accurate to ${\cal O}(\varepsilon^4)$. The success of the heuristic formula in 
\cite{JA} appeared to be the success of a mysterious rival to standard
adiabatic theory; but the mysterious rival has now been revealed as standard
adiabatic theory in a thin (but evidently useful) disguise.

Comparison with most other treatments of the quantum vector Kepler problem
is not relevant, because they examine predominantly cases of
small $J$, where the adiabatic method does not perform well. Ref.~\cite
{Green} provides numerical results in only one case where $\varepsilon $ is
really small, namely the exactly solvable case $s=\hbar /2$ and $J=11\hbar
/2 $. The reported results of the adiabatic calculation in \cite{Green} show
errors of order $\varepsilon ^{4}$ in this case. As far we understand, it is
only fortuitous that our results are actually better than this; but perhaps
some advantage is due to the fact that the adiabatic technique used in \cite
{Green} is uncontrolled, whereas ours is a systematic expansion in the small
parameter $\varepsilon $.  The effective potential (for the bound channel)
used in \cite{Green} is, in our units, 
\begin{equation}
U_{-}={\frac{1+\varepsilon ^{2}}{2r^{2}}}-\sqrt{r^{-2}+\varepsilon ^{2}r^{-4}}\;,
\label{HGreen}
\end{equation}
which is the exact lower eigenvalue of the the 2-by-2 potential matrix of 
(\ref{SE}) for $s=\hbar /2$. Expanding to ${\cal O}(\varepsilon ^{2})$ produces
the potential $H_{AD2}$, with its problematic $r^{-3}$ behaviour. As
we have seen, the canonical transformation to $H_{SS}$ effectively
removes this spurious singularity. Since (\ref{HGreen}) is also well-behaved
as $r\rightarrow 0$, and has the correct behaviour to ${\cal O}(\varepsilon
^{2})$ at large $r$, it is not surprising that it also gives results that
are good at ${\cal O}(\varepsilon ^{2})$. Obtaining still better results,
however, requires more than simply expanding $U_{-}$ to higher order in $\varepsilon $.

\section{VIII: Discussion}

Adiabatic methods are a remarkably powerful analytical tool. Their power
often does seem to be remarkable, in the sense that they frequently deliver
accuracy greater than one can naively expect.  The robustness of adiabatic
approximations has perhaps not yet been fully understood.
For example, an interesting issue that we have raised in this paper
concerns the possibility that the hierarchy of time scales, upon which
adiabatic methods depend, may break down within a localized region of phase
space. A classical system may then evolve outside this region for most of
its history, so that its behaviour can be closely approximated
adiabatically, but in a succession of intervals of `crisis', it may pass
briefly through the region where adiabatic methods fail to apply. During the
interval of crisis the fast degrees of freedom do not decouple from the
(nominally) slow ones, and so the precise state of the fast variables might
affect the slow variables in ways that could persist after the crisis. Thus 
the fast degrees of freedom, which within the adiabatic regime are
effectively hidden variables, could `emerge' during crises.

While this kind of emergence of hidden variables has (presumably) nothing to
do with quantum measurement, the classical phenomenon will have analogues in
quantum mechanics. The effective Hamiltonians given by standard adiabatic
techniques may be unreliable in some regions of Hilbert space. The result
may in principle be spurious simplicity, but it may also be spurious
complexity, in that the adiabatic effective Hamiltonians may have
singularities like the $r^{-3}$ potential of our $H_{AD2}$, which are
unphysical artifacts of applying the adiabatic formalism beyond its domain
of applicability.

In difficult cases of this kind, whether quantum or classical, a kind of
connection formula would seem to be needed, from which one could compute the
time and phase space location at which the system would exit the region of
adiabatic breakdown. In the problem studied in this paper, however, we have
seen that a co-ordinate change can eliminate a spurious singularity,
producing an effective Hamiltonian which does not seem to need a connection
formula. It would clearly be desirable to determine whether this is simply a
fortunate coincidence in this one case, or whether there may exist a general
method of identifying the best co-ordinates, and minimizing the damage done
by what we have called crisis regions.

\end{document}